\documentclass[prd,floatfix,twocolumn]{revtex4}
\usepackage{graphicx,ulem}
\usepackage{graphics,epsfig}
\usepackage{amssymb}
\usepackage{color}

\newcommand{\be}{\begin{equation}}
\newcommand{\ee}{\end{equation}}
\newcommand{\ba}{\begin{eqnarray}}
\newcommand{\ea}{\end{eqnarray}}
\newcommand{\bea}{\begin{eqnarray}}
\newcommand{\eea}{\end{eqnarray}}
\newcommand{\bc}{}

\def\hub{{\cal H}}

\begin{document}
\title{Nonlinear growth in  modified gravity theories of dark energy}

\author{Istvan Laszlo and Rachel Bean}
\affiliation{ Dept. of Astronomy, Space Sciences Building, Cornell University, Ithaca, NY, 14853, USA}

\begin{abstract}
Theoretical differences in the growth of structure offer the possibility that we might distinguish between modified gravity theories of dark energy and $\Lambda$CDM. A significant impediment to applying current and prospective large scale galaxy and weak lensing surveys to this problem is that, while the mildly nonlinear regime is important, there is a lack of numerical simulations of nonlinear growth in modified gravity theories. A major question exists as to whether existing analytical fits, created using simulations of  standard gravity, can be confidently applied. In this paper we address this, presenting results of N-body simulations of a variety of models where gravity is altered including  the Dvali, Gabadadze and Porrati model. We consider modifications that alter the Poisson equation and also consider the presence of anisotropic shear stress that alters how particles respond to the gravitational potential gradient. We establish how well  analytical fits of the matter power spectrum by Peacock and Dodds and Smith \textit{et al.}  are able to predict the nonlinear growth found in the simulations from $z=50$ up to today, and also consider implications for the weak lensing convergence power spectrum. We find that the analytical fits provide good agreement with the simulations, being within 1$\sigma$ of the simulation results for cases with and without anisotropic stress and for scale-dependent and independent modifications of the Poisson equation. No strong preference for either analytical fit is found.
\end{abstract}
\maketitle
\section{Introduction}\label{intro}
A diverse range of observations are showing consistent evidence for the acceleration of the universe's expansion, and the presence of dark energy, for example supernovae observations \cite{Riess:2004nr,Astier:2005qq,Riess:2006fw}, cosmic microwave background temperature and polarization fluctuations \cite{Spergel:2006hy,Hinshaw:2006ia,Jarosik:2006ib,Page:2006hz}, large scale structure surveys \cite{ Cole:2005sx,Tegmark:2006az} and baryon acoustic oscillations \cite{Eisenstein:2005su}. 

Interpretation of Einstein's cosmological constant as a vacuum energy requires the value to be fine-tuned to  far smaller than any theoretical expectation, e.g.\cite{Carroll:cosmoconst}), and has forced the exploration of alternative theoretical explanations. 
Since precision measurements of gravity only exist for scales $<10^{13}$ m (e.g.\cite{Adelberger:2003qq}), there is freedom to posit modifications of gravity acting on larger scales, such as those in \cite{Dvali:2000hr}. 

While cosmological observations of dark energy properties have so far focused on measurements of the homogeneous background density and astrophysical correlations in the linear regime, both theoretical and observational vistas are  now opening up that require a good understanding of the  growth of structure in the mildly nonlinear regime. Theoretically, measuring the growth of structure  might enable modified gravity theories to be distinguished from a standard cosmological scenario ($\Lambda$CDM) with a cosmological constant, $\Lambda$, and cold dark matter (CDM) \cite{Linder:astro-ph0701317,Amendola:2007rr, Dore:2007}. Observationally, the next generation of precision experiments will include weak lensing surveys, with several proposed large scale weak lensing experiments being developed in the coming decade, e.g. DUNE \cite{Refregier:2006dune}, JDEM/SNAP \cite{Albert:2005snap} and LSST \cite{Tyson:2006hs}.

Weak lensing is a potentially powerful probe of the late time evolution of the Universe, sensitive not only to the background expansion, but also able to give two-point and higher statistical correlations of the density field \cite{Takada:2003ef}, potentially in tomographic redshift slices \cite{Jain:2003tba,Bernstein:2003es}. 

Modified gravity models can introduce extrinsic anisotropic shear stresses (see e.g. \cite{Bean:2006up}) that modify the relationship between the weak lensing potential and the matter over-density that might be detectable by contrasting weak lensing with other large scale structure observations \cite{Zhang:2007aniso,Amendola:2007rr}.

 Many large scale structure statistics can be related to the underlying matter power spectrum, with nonlinear evolution at small scales. For standard general relativity, a typical approach is to use analytical fits based on N-body simulations of $\Lambda$CDM \cite{Peacock:pd,Smith:sp} and CDM with dark energy with an equation of state,$w$,$w$CDM \cite{Ma:1999dwa,McDonald:2005gz,Linder:2005hc} scenarios to apply the nonlinear correction to a linear power spectrum.
 Simulations of modified gravity models are for the most part lacking, however. With the exception of \cite{Stab:2006yuk,Shirata:2007qk}, analyses  often proceed by applying the $\Lambda$CDM based analytical nonlinear fits to modified linear power spectra, e.g. \cite{Shirata:2005yr,Amendola:2007rr}. Recently an analytical approach to estimating nonlinear growth in  modified gravity,  including those with anisotropic stress,  \cite{Hu:2007pj} was proposed and it was noted that there were currently no simulations against which to test the ansatz. 

In this work we directly address to what degree the nonlinear fits developed for standard gravity can be utilized for modified gravity theories, and establish whether the paucity of simulations in modified gravity theories can be excused. We first consider the applicability of standard gravity nonlinear fits to modified gravity theories in which just the Poisson equation is modified, considering the 5D gravity form considered by \cite{UAB:2001uab}, complementing the work of \cite{Stab:2006yuk,Shirata:2007qk} who considered nonlinear growth when a Yukawa-like gravitational coupling is introduced \cite{Sealfon:2005pl}. We then address the impact of anisotropic stress on nonlinear growth to assess if existing analytical nonlinear fits are adequate to model evolution in these scenarios. We consider the nonlinear growth in the Dvali, Gabadadze, and Porrati (DGP) 5D model \cite{Dvali:2000hr} and in toy models that contrast the effects of anisotropic stress with those of a modified Poisson equation. 

We first establish the framework for investigating modified gravity theories in Sec. \ref{theories}, and outline the specific models we consider with scale-independent and dependent modifications and the presence and absence of anisotropic shear.  The details of our simulations and implementation, including  the two standard analyitic fits are presented in Sec. \ref{method}. The approach to weak lensing is discussed in Sec. \ref{WLmethod}. The results showing dimensionless power spectra and the success of analytic fits are discussed in Sec. \ref{disc}. An overview of the conclusions and implications of our results is then presented in Sec. \ref{conc}. 

\section{Modified Gravity Theories}\label{theories}

We first outline the effect that the general modifications to gravity we study have on the perturbed Einstein's equations.
Following the notation of \cite{Ma:1995ey}, in the conformal Newtonian (or longitudinal) gauge, the metric is written as 
\ba
\label{eq:metric}
ds^2=a(\tau)^2\left[-(1+2\psi)d\tau^2+(1-2\phi)dx_jdx^j\right]
\ea
where $a$ is the expansion factor, $\tau$ is the conformal time, $x$ is the comoving coordinate ($j$=1,2,3 spatial directions) and $\phi$ and $\psi$ are the two gravitational metric perturbations. 

Einstein's equations relate the metric perturbations to fractional perturbations in density, $\delta_s \equiv\delta\rho_s/\rho_s$, peculiar velocity, $v_{(s)}$, and intrinsic shear $\sigma_s$ for a matter component ``$s$'',
\bea
k^2\phi + 3\hub(\dot\phi+\hub\psi) &=& -\frac{3\hub^2}{2}Q\sum_s\Omega_s\delta_s \label{eq1},
\\
k^2(\dot\phi+\hub\psi) &=& \frac{3\hub^2}{2}\sum_s(1+w_s)\Omega_s (ik^jv_{(s)j}),\ \ \ \label{eq2}
\\
\phi - \psi &=& \frac{9\hub^2}{2}\sum_s(1+w_s)\Omega_s\sigma_s + \sigma_0, \label{eq3}
\eea
where $\hub=\dot{a}/a$, $\Omega(a)$ is the fractional energy density, and $w(a)$ is the equation of state for the fluid. We have introduced the function $Q(k,a)$ as a modification in the relationship between the gravitational potentials and matter density in the $\delta T_{0}^{\ 0}$ equation, (\ref{eq1}), and $\sigma_0(k,a)$ as an extrinsic anisotropic stress in addition to the intrinsic anisotropic stresses from the matter components (predominantly radiation) in the equation for $\delta T_{i}^{\ j}$, $i\neq j$, (\ref{eq3}). For standard gravity $Q=1$ and $\sigma_0=0$.

Equations (\ref{eq1}) and (\ref{eq2}) combine to give
\bea
k^2\phi &=& -\frac{3\hub^{2}}{2}Q\sum_s\Omega_s\left(\delta_s + 3\hub(1+w_s) ik^j\frac{v_{(s)j}}{k^2}\right),
\\
&=& -\frac{3\hub^{2}}{2}Q\sum_s\Omega_s\Delta_s,
\eea
where $\Delta_s$ is a gauge invariant density variable defined in the rest frame of the matter components \cite{Kodama:1985bj}.

Density and velocity perturbations evolve according to the perturbed fluid equations which are unchanged by the gravitational modifications,
\bea
\dot\delta &=& -(1+w)(ik^jv_j-3\dot\phi)-3\hub(c_s^2-w)\delta,
\\
ik^j\dot v_j&=&-\left[ \hub(1-3w)+\frac{\dot{w}}{1+w}\right]ik^jv_j+\frac{c_s^2}{1+w}k^2\delta\nonumber \\
&& -k^2\sigma+k^2\psi, \ \ \ \label{veq}
\eea
where  $c_s^2$ is the sound speed for the fluid.

We will consider a Universe dominated by pressureless matter, $w_s=c_s^2=\sigma_s=0$,
and scenarios in which $\psi\sim\phi$, so that on subhorizon scales $|k^2\psi|\gg |3\hub\dot\phi|, |3\ddot\phi|$, and
\bea
\ddot\delta+\hub\dot\delta+k^2\psi&\approx& 0.
\eea
Using (\ref{veq}), we define the peculiar acceleration, $g$, 
 \bea
g_j\equiv \frac{1}{a}\frac{d}{d\tau}(av_j) &=& -ik_j\psi\label{pecacc}.
\eea

Following the notation of \cite{Amendola:2007rr}, we relate the anisotropic stress to $\phi$ through a function $\eta$, 
\bea
\eta \equiv \frac{ \sigma_0}{\phi}.
\eea
$Q$ and $\eta$ here are equivalent to $q$ and $\eta$ in \cite{Tsujikawa:2007gd}.

Making a subhorizon approximation,  and $\hub v /k \ll \delta$, assuming $v\lesssim \delta$, the modified Poisson equation and peculiar acceleration equations are
\bea
k^2\phi &=& -\frac{3\hub^{2}}{2}Q\Omega_m \delta,  \label{poisson}
\\
g_j &=& -ik_j(1+\eta)\phi. \label{pecacc2}
\eea
while the matter perturbation equation is
\bea
\ddot\delta&+&\hub\dot\delta-\frac{3\hub^{2}}{2}Q(1+\eta) \Omega_m\delta= 0. \label{mateq}
\eea
Note that, we can describe the evolution of $\delta$ in terms of the linear growth factor, $D$, with respect to some reference scale, $a_i$, $\delta(k,a) \equiv D(a)\delta(k,a_i)$ where $D$ is scale-independent for standard gravity, but could be scale-dependent if gravity is so modified.

We can relate the Fourier space modification to a real space interaction in the form of a Green's function,
\bea
\label{eq:integ}
\phi(\mathbf{r})&=&-G \rho_m(a) a^2 \int d^3 \mathbf{r}' \delta(\mathbf{r}')f(\mathbf{r}-\mathbf{r}'),\label{eq:Frphicomdimconvk}
\\
g(\mathbf{r}) &=& -\nabla\left[(1+\eta(\mathbf{r}))\phi(\mathbf{r})\right], \label{pecaccreal}
\eea
with $f(\mathbf{r})=1/|\mathbf{r}|$ recovering standard gravity. Using the convolution theorem we find,
\be
Q(k,a) = \frac{k^2}{4\pi}f(k,a).
\ee
 
The effect of modified gravity in weak lensing statistics is described in \cite{Schimd:2004nq} where they show that the weak lensing distortion is dependent upon the sum of the two gravitational potentials, $\Phi\equiv\phi+\psi$. As in \cite{Amendola:2007rr}, we introduce the parameter $\Sigma(Q,\eta)$ to describe the deviation of the weak lensing potential from standard gravity
\bea 
k^2\Phi&=&-3\hub^2\Sigma \Omega_m\delta,
\\
\Sigma&\equiv& \left(1+\frac{\eta}{2}\right)Q.
\eea
with $\Sigma=1$ for standard gravity.

\subsection{5D Gravity}\label{5D}
 We consider a model, motivated by 5-dimensional gravity theories in which gravity is Newtonian on small scales but modified on scales larger than a characteristic scale $r_s$ \cite{Gregory:2000jc,Binetruy:2000xv,UAB:2001uab}.
This model is characterized by the form 
\ba\label{eq:UABeq}
f(\mathbf{r})=\frac{1}{|\mathbf{r}|+{r^2 \over r_s}},
\ea
and
\ba
\nonumber Q(\mathbf{k},a)&=&\frac{k r_s}{2}\left[-2\left(\int_{k r_s}^\infty \frac{\cos(t)}{t} dt\right) \sin(k r_s)\right.\\
&&\left.+\cos(k r_s)\left(\pi-2\int_0^{k r_s}\frac{\sin(t)}{t} dt\right)\right],
\ea
with $\eta(k,a)=0$.

We are principally interested in the effect that modifications to gravity could have on the transition from linear to nonlinear regime, typically occurring over comoving scales $1-10 $ Mpc.  For our analysis, therefore, we consider evolution for values of the parameter $r_s$ of $20h^{-1}\ $ Mpc, $10h^{-1}\ $ Mpc, and $5h^{-1}\ $ Mpc, which alters the behavior in the relevant scales. We do not consider here smaller values of the modification which would alter behavior in the wholly nonlinear regime. We leave it for future study to assess whether such changes are well modeled by analytical fits describing the properties of collapsed halos.
 
\subsection{DGP}\label{DGP}
A physical model that serves as an excellent example of the effects of anisotropic shear is the Dvali, Gabadadze, and Porrati (DGP) model \cite{Dvali:2000hr} that is based on 5D gravity, wherein at some large scale, $r_c$ (comparable to the horizon scale), gravity is sensitive to the presence of an additional dimension.

The extra dimension alters the 4D background evolution to that described by the modified Friedmann equation,
\ba
\label{eq:HDGP}
H(a)={1 \over 2 r_c}+\sqrt{\left({1 \over 2 r_c}\right)^2+{8 \pi G \over 3}\rho_m(a)},
\ea
where $H=\hub/a$, with late time acceleration being triggered when the Universe's horizon $\sim r_c$.

The modification also alters the growth of fluctuations in density and motion of matter. As well as a modification to the Poisson equation as discussed in Sec. \ref{5D}, this model also results in an anisotropic shear such that the two potentials are given by \cite{Lue:2004rj,Koyama:2005kd,Maartens:2006dgp,Amendola:2007rr}:
\bea
\label{eq:poissonphidgp}
k^2 \phi&=&-\frac{3\hub^2}{2} \left(1-\frac{1}{3\beta}\right)\Omega_m\delta,
\\
\label{eq:poissonpsidgp}
k^2 \psi&=&-\frac{3\hub^2}{2} \left(1+\frac{1}{3\beta}\right)\Omega_m \delta,
\eea
where
\bea
\beta&\equiv&1-{2H^2(a) r_c^2 \over 2 H(a) r_c -1}.
\eea
In contrast to Sec. \ref{5D}, this gives a scale-independent modification to the Poisson equation,
\bea
Q(a)&=&1-{1 \over 3 \beta},
\eea
and nonnegligible anisotropic stress,
\bea
\eta(a)&=&{2 \over 3\beta-1},
\eea
and $\Sigma=1$.

For our analysis, with a background cosmology with Hubble constant, $H_0, = 70$ km s$^{-1}$ Mpc$^{-1}$, fractional matter density, $\Omega_{m},=0.3$, and consistent with the observational constraints found in \cite{Maartens:2006dgp} (\ref{eq:HDGP}), $r_c= 6.1$ Gpc.

\subsection{Twin toy models }\label{twin}
Finally, we consider a set of twin models that provide a simple way to further explore the effects of anisotropic stress on nonlinear growth. We consider two different modifications that both yield the same form for the weak lensing potential, with $\Sigma=1+\Sigma_0 a$, such that they reduce to standard gravity at early times and become modified at late times. This form of $\Sigma$ is equivalent to model GDE1 of \cite{Amendola:2007rr}. The twin models (``TM'') we study have two contrasting, simple forms in terms of $Q$ and $\eta$ :
\be
\begin{array}{lll}
\mathrm{TM \ 1}: &Q = 1, & \eta=2 \Sigma_0 a,
\end{array}
\ee
and
\be
\begin{array}{lll}
\mathrm{TM \ 2}: &Q = 1+\Sigma_0 a, & \eta=0.
\end{array}
\ee

 In TM1, the Poisson equation is the same as for standard gravity; however, the peculiar acceleration of the matter particles responding to the gradient of the potential is affected by the anisotropic stress. In TM2, in contrast, the peculiar acceleration is the same as for standard gravity but the gravitational potential at late times has a different relation to matter over/under densities. We consider values of $\Sigma_0 = \pm 0.008,\ \pm 0.016 $ consistent with 1 and 2$\sigma$  Fisher-matrix constraints for a  prospective DUNE-like weak lensing survey \cite{Amendola:2007rr}.
\section{N-Body Simulations}\label{method}
To obtain fully nonlinear results in each of the models, we obtain N-body simulations via a particle mesh(PM) code, taking as an initial form the code of \cite{Kly:1997co}. For scale-independent modifications we make simple modifications to the code, described in Sec. \ref{SG}. For scale-dependent modifications we have to alter the potential and motion calculations as described in Sec. \ref{scaledep}. 
\subsection{Standard Gravity and Scale-independent Modifications}\label{SG}
The PM code is reviewed in detail in \cite{Kly:1997co} but we provide some highlights in order to set the framework for discussing the modifications we make to the code.

PM codes operate by defining a simulation area as a box of size $L$ on a side, assuming it is closed so that we have periodic boundary conditions, subdividing it into a mesh or grid of $N^{3}$ cells (of size $L/N$ on a side), and defining all quantities on that mesh. The simulation is then initialized at some early redshift ($z_i$) and $N_{P}^{ \ 3}$ particles are placed according to model-dependent power spectra fits provided with the code (based on the cosmological parameters: the scalar spectral index $n_s$; the amplitude of fluctuations in 8$h^{-1}$ Mpc, $\sigma_8$;  the fractional density from curvature, $\Omega_K$, baryons, $\Omega_b$ and cold dark matter, $\Omega_{cdm}$; and the Hubble constant $H_0= 100 h$ km s$^{-1}$ Mpc$^{-1}$). The evolution is then carried out by advancing time in equal steps of the expansion factor, $a_{step}$. At each step in expansion factor the code determines a density in each cell, uses that density to compute the potential $\phi$ in each cell, and finally moves particles according to the gradient of the potential. 

\subsubsection{Defining the Density}
\begin{figure}[t!]
\scalebox{0.49}{\includegraphics{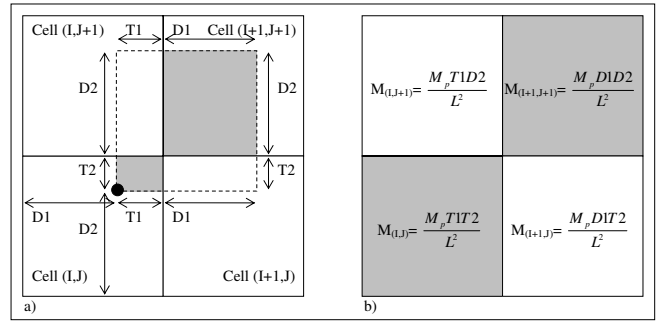}}
\caption{\label{fig:CIC}
A two dimensional description of cloud in cell density assignment. (a) The definition of the variables in relation to the particle's actual position. The particle is the black dot, but it is extended to be a square particle denoted by the dotted lines, thus it lies in four cells. The sides of the cells and the size of the particle square are $L=D1+T1=D2+T2$. (b) The resultant mass distribution in each cell. Note that the mass is not retained in the original particle's area, but rather smeared over the cell it occupies.}
\end{figure}

Defining the density can be done in a variety of ways; the code uses the cloud-in-cell scheme depicted in Fig. \ref{fig:CIC} wherein a particle is taken to be a cube with dimensions equal to that of the cells and with a corner positioned at the location of the particle. The particle contributes to each cell it extends into a mass equal to the particle's total mass weighted by the fraction of the particle's volume in the cell under consideration. Once the mass in each cell is determined it is effectively smeared over the entire cell.

\subsubsection{Obtaining the Potential}
For standard gravity, the code uses (\ref{poisson}) with $Q=1$, with the dimensionless variables of \cite{Kly:1997co}, $\tilde{r} \equiv r/x_0$ and $\tilde\phi\equiv \phi/(x_0 H_0)^2$ and writing $\delta\equiv \rho(x,a)/\bar{\rho}(a)-1$,
\be
\label{eq:Klypins}
\tilde{\nabla}^2 \tilde{\phi}={3 \over 2} {\Omega_{m,0} \over a}\delta.
\ee
To evaluate (\ref{eq:Klypins}), we use the discretized Poisson equation over cells, $n=0,N-1$. In one dimension, the discrete Laplacian is given by
\be
\label{eq:1d2ndderiva}
\nabla^2 \phi_{n} \approx \phi_{n+1}+\phi_{n-1}-2\phi_{n}.
\ee
Defining the discrete Fourier transform,
\be
\label{eq:phika}
\tilde{\phi}_{k}=\Sigma_{n=0}^{N-1}\phi_{n}e^{i 2 \pi n k/N},
\ee
the  discretized Poisson equation is
\be
\label{eq:phiexpa}
\nabla^2 \tilde{\phi}_{k}=\tilde{\phi}_{k} \times 2\left[\cos\left({2 \pi k \over N}\right)-1\right]. 
\ee
Generalizing to three dimensions one obtains the `7-point crest template',
\ba
\label{eq:sevenptcresta}
\nonumber \nabla^2 \phi_{i,j,k} \approx && \phi_{i+1,j,k}+\phi_{i-1,j,k}+\phi_{i,j+1,k}+\phi_{i,j-1,k}\\
&+&\phi_{i,j,k+1}+\phi_{i,j,k-1}-6\phi_{i,j,k}, \ \ \ \ \ \ \ 
\ea
with
\be
\label{eq:phiexp3da}
\nonumber \nabla^2 \tilde{\phi}_{k}=\tilde{\phi}_{k} \times G_k,
\ee
where $G_k$ is given by 
\be
\label{eq:gka}
G_k=2\left[\cos\left({2 \pi k_x \over N}\right)+\cos\left({2 \pi k_y \over N}\right)+\cos\left({2 \pi k_z \over N}\right)-3\right].
\ee
Combining  (\ref{eq:phiexp3da}) and (\ref{eq:Klypins}) the Poisson equation used in the code is,
\ba
\label{eq:pmcode}
\tilde{\phi}_{k} = {3 \over 2} {\Omega_{m,0} \over a \ G_k}\delta.
\ea
The code calculates  $\delta(\bf{r})$, Fourier transforms to $\delta(\bf{k})$, divides by $G_{k}$ and then transforms back to real space to obtain $\phi(i,j,k)$. 

In the case of scale-indepedent modifications (\ref{eq:pmcode}) is purely modified by 
\bea
G_{k,alt}(k,a) \equiv \frac{G_k}{Q(a)}.
\eea
\subsubsection{Advancing the Particles}
Once we have the potential $\phi$ we  advance the particles according to (\ref{pecaccreal}). In standard gravity,  component wise on the grid we have only to compute 
\bea
\nonumber g_x&=&-(\phi_{i+1,j,k}-\phi_{i-1,j,k})/2\\
\nonumber g_y&=&-(\phi_{i,j+1,k}-\phi_{i,j-1,k})/2\\
\nonumber g_z&=&-(\phi_{i,j,k+1}-\phi_{i,j,k-1})/2.
\eea
In the presence of anisotropic stress modifications,
\bea
g_{j,alt} &=& [1+\eta(a)]g_{j}.
\eea
\subsection{Scale-dependent Modifications}\label{scaledep}
 In order to incorporate the scale-dependent modifications to gravity we follow the convolution approach in (\ref{eq:Frphicomdimconvk}). To do this we multiply by $f(\mathbf{k},a)$ at each step in $a$ rather than $1/G_k$. 
\subsubsection{Defining the Radius $\mathbf{r}$ for $g(\mathbf{r})$}\label{defr}
In scale-dependent theories, by definition, we now convolve with functions involving the actual scale $r$, and we must therefore define explicitly a radius on the grid. The mass is smeared over the entire cell it lies in, so that the distances simply become those between cells. The Fourier transforms involve periodic boundary conditions, so we define the radius for one origin at $(0,0,0)$, and wrap the radius around the grid. Since the code uses the dimensionless radii to compute the function we have called $f(\mathbf{\tilde{r}})$, the cell indices can be used to construct the radius and we define $\tilde{r}_i$ to be the index of the relevant cell in the $i^{th}$ direction ($i$=1,3).

The periodic boundary conditions require a change to the basic prescription presented above, namely to include the periodic boundary condition we must set up a 1D radius of the form $[0,\ 1, \ 2, \ ..., \ N/2-1, \ N/2, \ N/2-1, \ ...,  \ 2, \ 1]$ where $N$ is the number of cells making up the grid in a dimension. Thus, the radii in the i-th dimension can be defined as
\ba
\tilde{r}_i(n)=\left\{ \begin{array}{cc}
n & \ \ n\le N_i/2\\
N_i-n & \ \ n > N_i/2 \end{array} \right.
\ea
The final 3-dimensional radius, $\tilde{r}$, is computed trivially as 
\ba
\tilde{r}^2=\sum_{i=1}^3 \tilde{r}_i^2.
\ea

There remains one final subtlety in computing the radius. Since $\tilde{r}[1,1,1]=0$, division by $\tilde{r}$ requires us to make a change to avoid infinities. To avoid these singularities we take the standard approach of `softening' $\tilde{r}$ (e. g. \cite{Hockney:pm}), that is adding a small non-zero term to all the values of $\tilde{r}$ used in operations that would give a singularity. For instance if we consider $g(\tilde{r})= 1/\tilde{r}$ we instead compute $g(\tilde{r})= (\tilde{r}^2+\epsilon^2)^{-1/2}$. Note that, for consistency, all values of r in the division are softened, not only the actual one that gives a singularity ($\tilde{r}[1,1,1]$). Further, note in the case of well defined modifications, e.g.  $g(\tilde{r})=e^{-\tilde{r}} /\tilde{r}$ the exponent need not be softened, so that we compute $e^{-\tilde{r}}(\tilde{r}^2+\epsilon^2)^{-1/2}$. 

\subsection {Obtaining Analytic Spectra}\label{analytic}

We compare the nonlinear spectra from simulations to predicted spectra from analytical mappings of linear power spectra using the Peacock and Dodds (PD) fit \cite{Peacock:pd} and the Smith \textit{et al.} fit (SP) \cite{Smith:sp}.

\subsubsection{Analytical Linear Spectra}\label{linearfits}
\begin{figure}[t!]
\scalebox{0.90}{\includegraphics{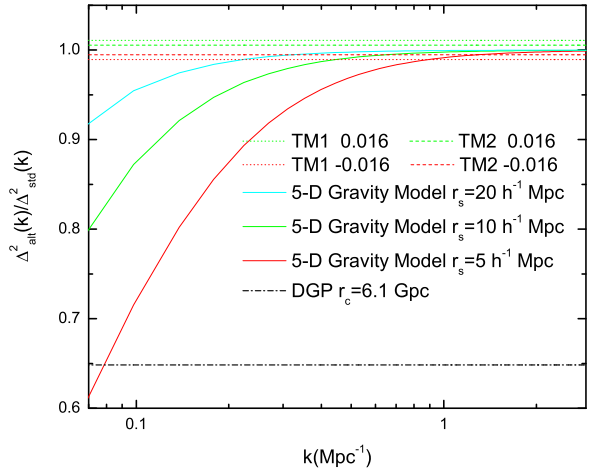}}
\caption{\label{fig:LinearPower}
The ratio of the linear power spectrum in the modified theories to that for standard gravity for the models discussed in Sec. \ref{theories}: the 5-D gravity model of Uzan and Bernadeau (solid line), TM1 (dotted line),  TM2(dashed line) and DGP (dotted-dashed line).}
\end{figure}

We evolve a linear $\Lambda$CDM power spectrum obtained with CAMB \cite{Lewis:2002ah} (that includes effects from baryon photon coupling at early times) forward in time using the modified equation for the growth of the over-density (\ref{mateq}). We start at an epoch, we choose $z_i=50$, at which the modification scale is large compared to the physical horizon, so that standard gravity is effectively recovered on the relevant scales, and evolve the density perturbations through the modified gravity era to today. In Fig. \ref{fig:LinearPower} we show the linear power spectra for the models discussed in the paper.

\subsubsection{Analytical Non-Linear Fits}\label{fits}

We briefly review the physical ingredients of the PD \cite{Peacock:pd} and SP \cite{Smith:sp} analytical fits against which we compare the simulations.

The PD fit is based on the assumption of stable clustering \cite{1977ApJS...34..425D}, the hypothesis that the correlation function on scales smaller than those of virialized structures decouple from the expansion. The fit utilizes a linear to nonlinear mapping  proposed by Hamilton \textit{et al.} (HKML) \cite{Hamilton:1991hk} 
\bea
k_L&=&[1+\Delta^2_{NL}(k_{NL})]^{-1/3}k_{NL}.
\eea
derived from the spherical collapse model. Peacock and Dodds generalized the HKML method to estimate the resulting nonlinear power spectrum through a universal scaling relation, $f_{NL}$,
\bea
\Delta^2_{NL}(k_{NL})&=&f_{NL}[\Delta^2_L(k_{L})],
\\
f_{NL}(x) &=& x\left[\frac{1+B\beta x+[Ax]^{\alpha\beta}}{1+([Ax]^{\alpha}g^3(a)/[Vx^{1/2}])^{\beta}}\right]^{1/\beta},\ \ \ \ \ 
\eea
where $g\equiv D(a)/a$.  The fitting function $f_{NL}$ tends to $f_{NL}(x)=x$ in the linear limit, $x\ll 1$, and $f_{NL}(x)=V g^{-3}(\Omega_m)x^{3/2}$ in the small scale, stable clustering limit, $x\gg 1$. There are five free parameters fit from N-body simulations in standard gravity as functions of the linear spectral index $n_{eff} = d\ln P_\delta/d\ln k(k=k_L/2)$: $A$ and $\alpha$ parameterize the power law in the quasilinear, large scale regime, $V$ parameterizes the amplitude of the $f_{NL}(x)$ in the stable clustering limit, $B$ describes the second order deviation from linear growth and $\beta$ softens the transition between the linear and fully virialized regimes. The cosmological model only enters into the fit through $g$, consistent with the Zel'dovich approximation in which the final positions of particles are obtained by extrapolating their initial comoving displacements, $q$,  using the linear growth factor, $x(a,t) =a(t)\left[ q + D(a) \nabla \psi(q)\right]$.

The quality of the PD fit is founded on the broad applicability of the Zel'dovich approximation. However, with a scale-dependent modification of gravity, or the introduction of a difference between $\phi$ and $\psi$ it is not clear {\it a priori} how well the Zel'dovich approximation will apply, and if applicable, whether the numerical values of the coefficients will remain the same as those for standard gravity. That is, with scale-dependent modifications the possibility for shell crossings arises which causes a break down of the Zel'dovich approximation.   

Looking at the functional form of the fit, in particular, we can consider three regimes to make predictions, namely, the large and small scale limits and a transition regime. Large scales which remain linear or quasilinear should be well described by the existing fit. On these scales the Zel'dovich approximation should hold and using a linear growth factor for $g$ is acceptable. Similarly $\alpha$ and $\beta$ might be expected to adapt to the changed input power via their spectral index dependence, since in linear scales essentially all the information is contained in the amplitude and spectral index of the power spectrum.

The mildly nonlinear or transition regime, where we directly compare results, is particularly of interest in applying the fits. Scale-dependent modifications introduce an extra degree of freedom to growth in the model, a scale dependency that could also affect the shape and scale of the smoothing function interpolating between the linear and nonlinear asymptotic behaviors, essentially requiring corrections to $\beta$. For example, a scale- or time-dependent modification to Poisson's equation could alter the critical over-density required for nonlinear collapse, thus altering the details of the transition from linear to nonlinear regimes.

Small scale modifications to gravity, which we do not consider here, could well lead to alterations in the correlation function of the collapsed structures, in particular, changes to the value of the virialized normailzation $V$. One might expect the application of the linear growth factor in the fit to be less effective even if including the linear scale dependency $g(a)\rightarrow g(k,a)$.  Relevant to our analysis is the fact that the stable clustering approximation does not account for merging and accretion of halos and hence does not address how modifications to gravity may alter these physical processes. We discuss this in the context of the SP fit below.

The SP fit arises from a different approach based on the ``halo model'' \cite{Seljak:2000,Peacock:2000} in which the continuous accretion of matter and merging of halos is accounted for, deviating away from the stable clustering approximation. In this scenario, the power spectrum of matter $\Delta_{NL}^{2} = \Delta_{Q}^{2}+\Delta_H^2$ is described on large scales by the correlations between different halos represented by a quasilinear term, $ \Delta_{Q}^2(k)$, and on small scales by a halo term, $\Delta_{H}^2(k) $, that accounts for power from the self-correlation of halos.   In the fit, the two terms are phenomenologically selected functions of $y\equiv k/k_\sigma$, where the scale $k_\sigma(a)$ becomes nonlinear at scale factor $a(t)$,
\bea
\Delta_Q^{2}(k) &=& \Delta_L^{2} (k)\left[\frac{\left(1+\Delta_L^{2}(k)\right)^{\beta_n}}{1+\alpha_n\Delta_L^2(k)}\right]\exp\left(-\frac{y}{4}+\frac{y^2}{8}\right), \ \ \ \ \ \ \ \ \\
\nonumber \Delta_H^{2}(k) &=& \frac{a_n y^{3f_1(\Omega_m)}}{1+b_ny^{f_2(\Omega_m)}+(c_nf_3(\Omega_m)y)^{3-\gamma_n}}\times
\\ && \frac{1}{1+\mu_ny^{-1}+\nu_ny^{-2}}.
\eea
$k_\sigma$ is determined by the standard error of the linear density field, 
\bea
\sigma(k_\sigma^{-1},a)&\equiv& 1,
\\
\sigma(R,a) &\equiv& \int \Delta_L^2(k,a)\exp(-k^2R^2)d\ln k. 
\eea
The eight coefficients \{$\alpha_n,\beta_n,\gamma_n,\mu_n,\nu_n,a_n,b_n,c_n$\}, fit with spectral index-dependent functions, and three $\Omega_m$-dependent functions, $f_1$, $f_2$, and $f_3$, are empirically matched to standard gravity simulations.

In the large scale limit, the quasilinear term dominates and the use of the spectral index dependent $\alpha_n$ and $\beta_n$ suggest the fit will adapt well to a modification on linear scales, such as those considered here.

On small scales, just as in the PD case, there are issues with the numerical fitting functions in the halo self-correlation term; the correlation coefficient in a virialized halo could be modified for the various modified gravity scenarios. The functions $f_1$, $f_2$, and $f_3$, which in standard gravity are purely functions of $\Omega_m$, would be expected to alter to account for the modification;  this in turn could well be expected to change  $a_n$, $b_n$, $c_n$. 

Also as in the PD case, the interpolation from linear to nonlinear regimes, from large to small scales, could be altered as the modifications could alter the critical over-density required for nonlinear collapse. In particular the form of $a_n$ and to some extent $\beta_n$ may be expected to require changes as these serve to determine the relative importance of the halo-halo and self-correlation terms.

It is in light of these considerations that we study whether these analytic fits can readily describe modified gravity scenarios, with scale-dependent or independent modifications to the Poisson equation, and/or scale-independent anisotropic shear. 
 
\section{Obtaining Weak Lensing Spectra}\label{WLmethod}
Modified gravity theories can impact weak lensing convergence power spectrum in addition to the matter power spectrum, thus we study the impact of our models on both. 
In standard gravity, the power spectrum of the convergence is given by
\ba
\label{eq:convergence}
P_\kappa(l)&=&\frac{9 \Omega_{m,o}^2 H_0^4}{4 c^4} \frac{1}{4}\int_0^{\chi_s} \frac{g^2(\chi)}{a^2 \chi^2} P_\delta\left(\frac{l}{\chi}\right) d\chi,
\ea
where $P_\delta$ is the matter power spectrum and $g(\chi)$ is a weighting function that can be related to the comoving distance $\chi$ and the distribution of background or source galaxies, $W_s(\chi)$
\ba
g(\chi)=2 \chi\int_\chi^{\chi_s} \frac{\chi'-\chi}{\chi'}W_s(\chi') d\chi'.
\ea  
We assume a a simple delta function distribution of sources at $z_s=1$, so,
\ba
g(\chi)=2 \chi \frac{\chi_s-\chi}{\chi_s}.
\ea  

The convergence power spectrum is then
\ba
P_\kappa(l) =\frac{9 \Omega_{m,o}^2 H_0^4}{4 c^3}\int^1_{a_s} \frac{W^2(\chi,\chi_s)}{a^4 H(a) \chi(a)^2} P_\delta\left(\frac{l}{\chi(a)}\right) da
\label{kappaint}
\ea
with
\ba
W(\chi,\chi_s)\equiv\chi \left(\frac{\chi_s-\chi}{\chi_s}\right).
\ea
Where we have used the fact that the comoving distance, $\chi$, is equal to the (comoving) angular diameter distance for a flat Universe so that
\ba
\chi(a)=\int_a^1 \frac{c da'}{a'^2 H(a')}.
\ea
Gravitational modifications $Q\neq 1$ and/or $\eta\neq 0$,  will act to modify $P_{\delta}(k,a)$. In addition, $Q\neq 1$ and/or $\eta\neq0$ will modify how the convergence spectrum is related to $P_{\delta}$ (see for example \cite{Amendola:2007rr}), resulting in
\ba
\label{eq:WLintgeneral}
\nonumber P_\kappa(l) &=&\frac{9 \Omega_{m,o}^2 H_0^4}{4 c^3}\times \\
&&\int^1_{a_s} \frac{W^2(\chi,\chi_s)(1+\eta/2)^2Q^2}{a^4 H(a) \chi(a)^2} P_\delta\left(\frac{l}{\chi(a)}\right) da,\ \ \ \ \ \ 
\ea
where the evolution of $P_\delta$ is also affected by $Q$ and $\eta$.

$P_\delta$ can be obtained from either the PM simulations or the analytic fits described in Sec. \ref{analytic}.
To actually evaluate the integral, we discretize it, binning by expansion factor.

When considering the N-body code derived $P_\delta$, we have to account for the fact that the simulation only probes a range of $k$, yet for any given $l$, $k=l/\chi(a)$ can can lie outside this range at some redshift, $z_s>z>0$. For the $l$ range we consider, the range of $k$ needed is virtually all given by the N-body simulation. Outside this range, on large scales the power spectrum is well approximated by the linear spectrum; at smaller scales we find that the analytic SP and PD predictions for the modified gravity spectra are within the $1\sigma$ errors at the edges of the range of $k$ provided by the simulations, so we pad the simulated spectra with the nonlinear analytic fits to modified gravity linear power spectrum. 

\section{Results}\label{disc}

\subsection{Parameters for Simulations}\label{disc:Params}

For the PM code parameters we take $N=256,\ N_P=128,\ L=100 h^{-1}$ Mpc,$\ \epsilon=0.1,\ z_i=50,\ a_{step}=0.002$ and for our cosmological model we take $n_s=1,\ \sigma_8=1,\ \Omega_K=0,\ \Omega_m=0.3,\ \Omega_b=0.026,\ \Omega_{cdm}=0.274$, and $h=0.7$. The resulting simulations measure scales $0.1\lesssim k\lesssim 1$ Mpc$^{-1}$.  The specific choice of initial redshift is not important other than to ensure that it is early enough that nonlinear corrections are negligible.

The box size and number of cells play into spatial resolution of the simulation, and are chosen to allow us to effectively probe the decade of $k$ in which the mildly nonlinear effects manifest themselves and from which we can extract a reasonable weak lensing spectrum for $l\sim200-1000$, a range relevant to upcoming experiments. The number of particles are chosen to ensure a sufficient particle resolution for the box size and number of cells used. 

The initial positions of the particles at $z_i$, are assigned by means of a random number generator consistent with the initial power spectrum. Depending on the seed used to initialize the random number generator, the resultant spectra may agree well with standard $\Lambda$CDM analytic fits with the same parameters or might over- or under-produce power, even in the original unaltered code of \cite{Kly:1997co}. We therefore run the simulations with 24 random seeds to get a good sample size and a more robust average. In order to weight the behavior of each simulation equally, we consider the modifications in terms of the ``average of the ratios" of the modified power spectrum to the standard gravity spectrum for the same seed, rather than the ``ratio of the averages" that would preferentially weight those simulations that over-produce power. 

For scale-dependent modified gravity, we find the Numerical Recipes routine \cite{Press:nr} for the Fourier transform, though slightly more time consuming, is more stable than the one provided in the original code. In the case of scale-independent modifications and standard gravity, both algorithms produce identical results. The softening parameter value  used for the scale-dependent modification is much smaller than the smallest separation in the code and provides agreement with standard gravity from analytic predictions and standard gravity simulations with the code at least at the level or better than the unmodified Klypin code. 

\begin{figure}[t]\scalebox{0.90}{\includegraphics{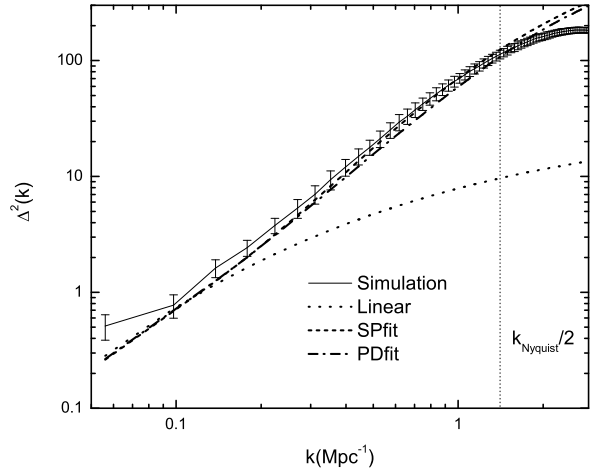}}
\caption{\label{fig:del2}
Dimensionless matter power spectrum, $\Delta^2(k)\equiv k^3 P_{\delta}(k)/2\pi^2$, for standard gravity. The full line and errors bars show the average power spectrum and standard deviation for 24 simulations. The vertical dotted line represents $k_{Nyquist}/2$, which is a conservative estimate for the largest $k$ at which we can believe the simulation results as in \cite{Stab:2006yuk}. The PD (dot-dashed line) and SP (dashed line) analytical fits are also shown.
}
\end{figure}

\begin{figure}[t!]
\scalebox{0.9}{\includegraphics{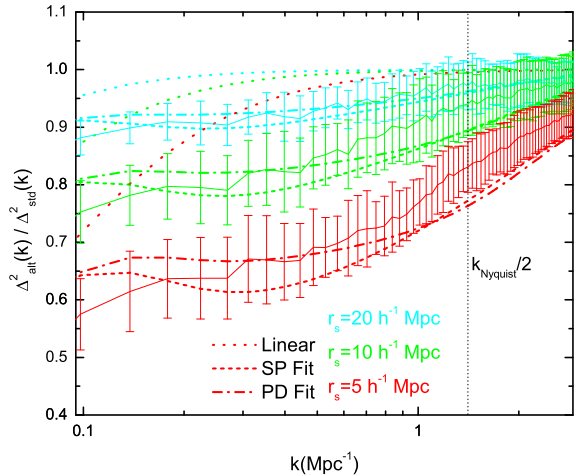}}
\caption{\label{fig:UABratios}
Ratios of the $z=0$ dimensionless matter power spectrum in the modified gravity model to that for standard gravity,  for the 5D gravity model described in Sec. \ref{5D} for $r_{s}=20h^{-1}$ Mpc (top, blue), $10h^{-1}$ Mpc (middle, green) and $5h^{-1}$ Mpc (bottom, red). The full line and errors bars show the average of the ratios and standard deviation for 24 simulations. The vertical dotted line represents $k_{Nyquist}/2$, which is a conservative estimate for the largest $k$ at which we can believe the simulation results as in \cite{Stab:2006yuk}. The PD (dot-dashed line) and SP (dashed line) analytical fits agree with simulations to within 1$\sigma$ for each $r_s$, in the region of interest, $k=0.1$ to $1$ Mpc$^{-1}$. 
}
\end{figure}

\begin{figure}[t!]
\scalebox{0.90}{\includegraphics{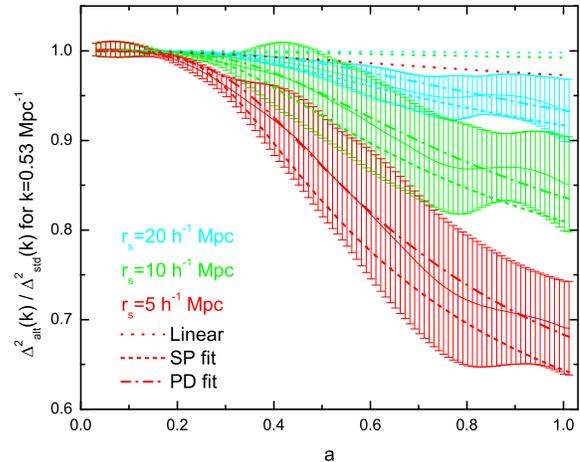}}
\caption{\label{fig:UABovera}
The ratios of the dimensionless matter power spectrum in modified to standard gravity, $\Delta_{alt}^2(k)/\Delta^2_{std}(k)$ as a function of redshift $50\le z \le 0$ for $k=0.53$ Mpc$^{-1}$. The color coding and lines styles are as in Fig. \ref{fig:UABratios}. The dotted lines show the ratios of the associated linear spectra. Note that the evolution is well tracked by the analytical fits, with both lying within 1$\sigma$ for the simulations. At late times the SP fit drifts to around, or just over, the 1$\sigma$ error.
}
\end{figure}

 In Fig. \ref{fig:del2} we show the results of the 24 simulations of standard gravity against the SP and PD fits, in order to demonstrate the fiducial model against which the modified gravity simulations are compared. The simulations are consistent with the analytical fits in the range $0.1$ Mpc$^{-1}\lesssim k\lesssim 1$ Mpc$^{-1}$. A conservative estimate for the largest $k$ at which we can believe the simulation results  are reasonable is $k_{Nyquist}/2 $ \cite{Stab:2006yuk}, which for our simulations is $1.4$ Mpc$^{-1}$. We consider the simulations to be valid only up to $k_{Nyquist}/2$, rather than up to $k_{Nyquist}$ as this more conservative limit represents a regime in which standard gravity simulations and fits agree to within 1.5 times the standard error in the simulation, in comparison to 10 (for the PD fit) and 13 (for the SP fit) times the standard error at $k_{Nyquist}$.

For the model parameterizations we consider, we find that the linear scales used to generate the nonlinear $k$ in the range $0.1-1.4$ Mpc$^{-1}$ lie in the range $k\sim0.07-0.5 $ Mpc$^{-1}$.

\subsection{Simulation and Analytic Fit Results}
\subsubsection{5D Gravity Model}

The ratio of the dimensionless power spectrum today for the 5D gravity model discussed in Sec. \ref{5D} to standard gravity,  is shown in Fig. \ref{fig:UABratios}.  We find that the simulations are consistent with the PD \cite{Peacock:pd} fit at the $1\sigma$ level. This is consistent with the results of \cite{Stab:2006yuk} for a Yukawa type modification (that, like the modification we consider here, is a scale-dependent modification). The SP fits are slightly less consistent with the numerical predictions, however, still lie within 1$\sigma$ of the simulation mean. We, therefore, find no statistical basis for preferring PD over the SP \cite{Smith:sp} fit of $P_{\delta}(z=0)$.

To consider the suitability of the analytic fitting functions when applied to  weak lensing, it is insufficient to purely consider their agreement with predictions today;  the entire evolution must be tracked between the redshift of the lensed source and today, as weak lensing integrates $P(k,a)$ over the expansion factor $a$, c.f. (\ref{eq:WLintgeneral}). We, therefore, track the redshift history of the nonlinear  evolution, and the comparison with the analytical fits, as shown in Fig. \ref{fig:UABovera}. We find both fits lie within 1$\sigma$ though after $a\sim0.75$ the SP results are just encompassed by the 1$\sigma$ errors.

\begin{figure*}[t!]
\centerline{\hbox{\hspace{0.0in}
\scalebox{0.90}{\includegraphics{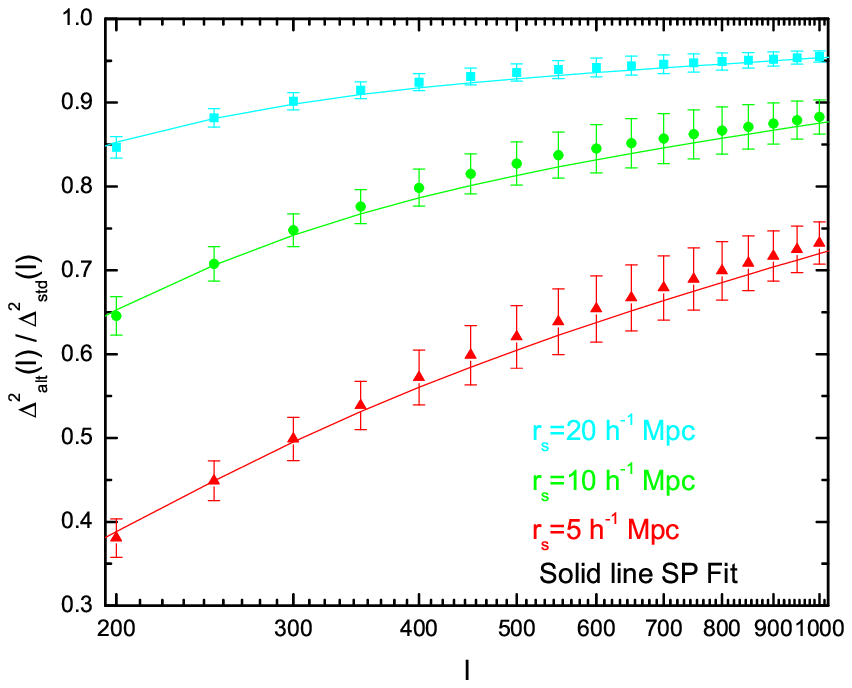}}
\scalebox{0.90}{\includegraphics{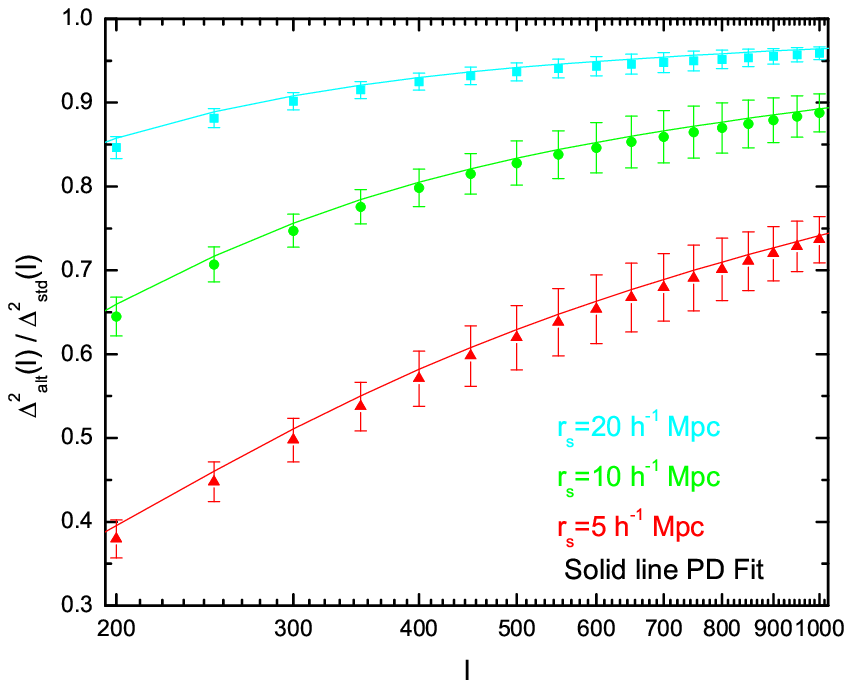}}}}
\caption{\label{fig:UABWL24}
The ratio of the weak lensing dimensionless convergence power spectrum,  $\Delta^2(l)\equiv l^2 P_\kappa(l)/2\pi$, for a $\delta$ function lensing source at $z_s=1$, as a function of multipole, $l$,  for the 5D gravity model in Sec. \ref{5D} to that in standard gravity in comparison to the SP fit (left-hand panel) and PD fit (right-hand panel).  The points and errors are the average and standard deviation of the ratios the 24 simulations. The predicted spectra from the analytical fits (full lines) are wholly consistent with the simulations for all 3 modified gravity models with $r_s=20 h^{-1}$ Mpc (top, blue),  $10 h^{-1}$ Mpc (middle, green) and $5 h^{-1}$ Mpc (bottom, red).}
\end{figure*}

The ratios of the modified gravity weak lensing spectra to those of standard gravity are well recovered by the PD and SP fits, as shown in Fig. \ref{fig:UABWL24}. The ratios of the weak lensing convergence spectra are slightly less sensitive to the exact form of the modification  than the matter power spectra, for two reasons. First,  the integral in (\ref{eq:WLintgeneral}) is mostly weighted towards integrand values at early times when the analytical fits are in very strong agreement with the simulations. Thus, for instance, the late-time transition of the SP fit to the outer regions of the 1-sigma level is not so significant to the convergence power as it is to the final matter power spectrum. Secondly,  we ``pad'' the spectrum  at $k$ values outside the simulated range with the analytical fit, in order to evaluate $l=k\chi(a)$ in (\ref{eq:WLintgeneral}). This is mitigated (as discussed in Sec. \ref{WLmethod}) by the similarity of the fits and the code spectra at the edges of our range of $k$ and the fact that the contribution from padded $k$ values is small in comparison to those drawn from the simulated range:  for $l=200$, $P_{\kappa}$ is padded with the nonlinear analytical spectrum at $a>0.955$, which corresponds to 1.3\% of $P_{\kappa}$ for standard gravity; for $l=1000$ the padding is required for $0.8<a<1.0$ which contributes to 14\% of the value of $P_{\kappa}$.

\subsubsection{DGP}
The effects of nonlinear growth in DGP models are of great interest in establishing observational distinctions between this model and standard $\Lambda$CDM at cosmological scales, in  \cite{Amendola:2007rr} the nonlinear power spectrum was estimated using the Smith \textit{et al.} analytical fit, while in \cite{Hu:2007pj} an analytical ansatz is applied. Both the DGP model and the model in Sec. \ref{5D} are motivated by 5D modifications to gravity. The difference between DGP and that model is that DGP not only modifies the Poisson equation but also the peculiar acceleration through the presence of an anisotropic stress. 

\begin{figure}[t!]\scalebox{0.90}{\includegraphics{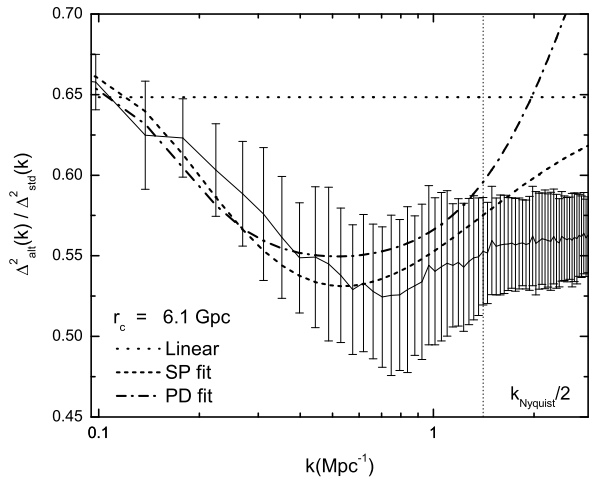}}
\caption{\label{fig:DGPratios24}
Ratios of the matter power spectrum in the DGP model with $r_c=6.1$ Gpc$^{-1}$ to that in standard gravity; both models have $H_0=70$ km s$^{-1}$ Mpc$^{-1}$ and $\Omega_{m}=0.3$. The full line is the average of the 24 realizations and errors represent the standard deviation of the simulations. The SP (dashed line) and PD (dotted-dashed line) analytic fits are in good agreement over the scales measured by the simulation, $k=0.1$ to $1$ Mpc$^{-1}$. The linear power spectrum ratio is shown by the dotted line.
}
\end{figure}

\begin{figure}[t!]
\scalebox{0.90}{\includegraphics{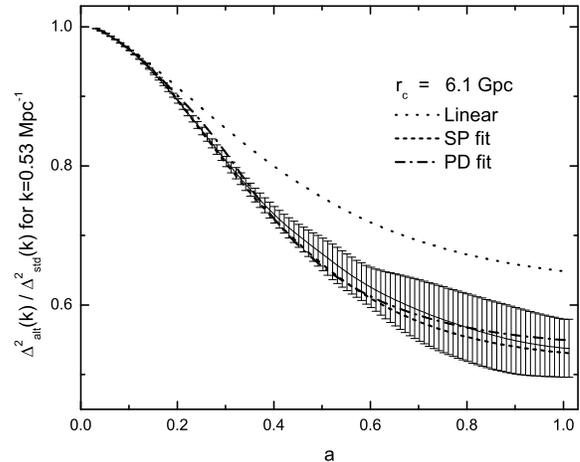}}
\caption{\label{fig:DGPovera24}
The evolution of the ratio of the DGP matter power spectrum to standard gravity for $k=0.53$ Mpc$^{-1}$ as a function of scale factor, $a$. The full line is the average of the 24 realizations and errors represent the standard deviation of the simulations.   The SP (dotted line) and PD (dotted-dashed line) analytic fits are good at  predicting the transition and development of nonlinear growth at all epochs.
 }
\end{figure}

For the arguably more complex DGP model, the SP and PD fits are both still in good  agreement with the N-body simulations at $a=1$, at the 1$\sigma$  level over the simulated scales, as shown in Fig. \ref{fig:DGPratios24}. This is also true over the course of the evolution as the modification from $\Lambda$CDM switches on, as shown in Fig. \ref{fig:DGPovera24}. 
 
 Note that we do not provide a weak lensing analysis in this model; as due to the change in $H$ [and hence in $\chi(a)$] evaluating $k= l /\chi(a)$ consistently results in a need for much smaller scales, i.e. $k\gtrsim 4.6$ Mpc$^{-1}$ for the range of $l$'s we have considered. We thus restrict our discussion of DGP to matter power spectra and their evolution. 
 
 Even though $r_c$ is chosen to be in close agreement with the background evolution of our fiducial cosmological model, and has essentially degenerate evolution at early times, the DGP model shows marked deviation from standard gravity at late times. We note that the suppression of the nonlinear power spectrum shown with respect to standard gravity for the PD and SP fits and N-body simulations  is qualitatively similar to that shown with the ansatz of \cite{Hu:2007pj}, although we leave a quantitative assessment of the ansatz to future work.

\subsubsection{Twin Toy Models}

\begin{figure*}[h!]
\centerline{\hbox{\hspace{0.0in}
\scalebox{0.90}{\includegraphics{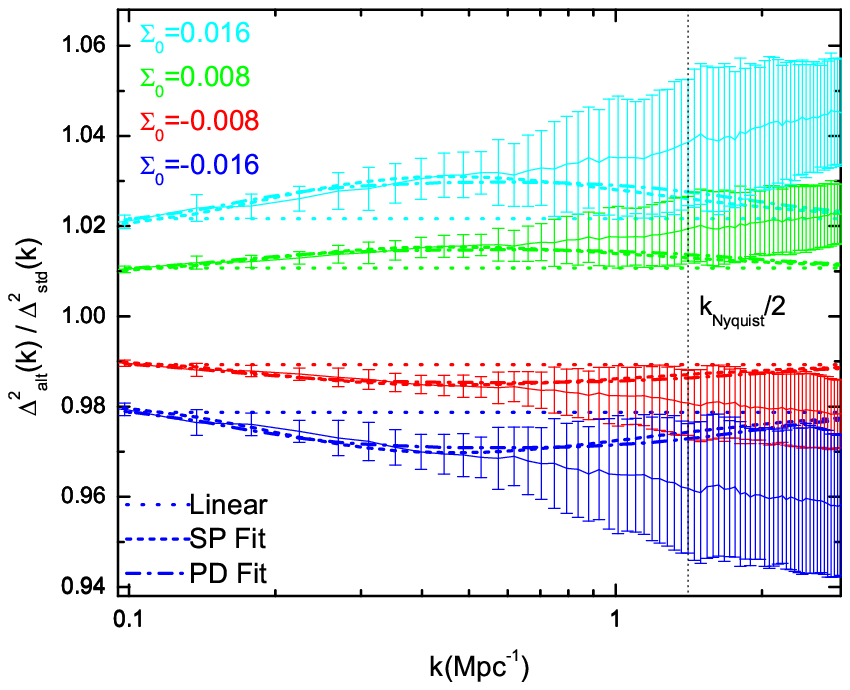}}
\scalebox{0.90}{\includegraphics{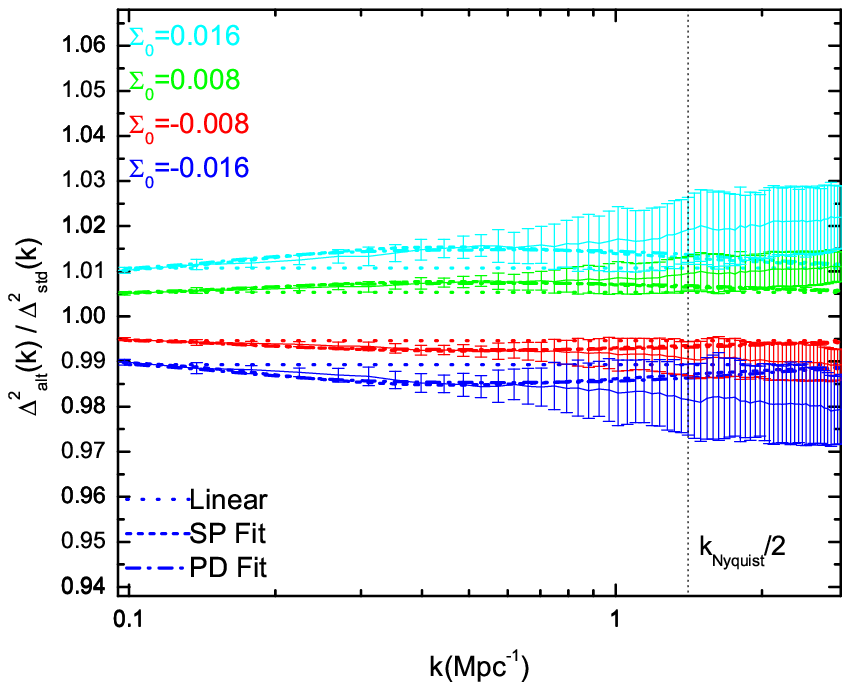}}}}
\caption{\label{fig:TMratios}
The ratios of matter power spectra at $a=1$ for modified gravity to standard gravity in the TM1(left panel) and TM2 (right panel) models for $\Sigma_0 = -0.016$ (dark blue, bottom), -0.008 (red) ,0.008 (green) and 0.016 (light blue,top) as a function of scale, $k$.  As in earlier figures, the full line represents the average of the 24 simulations, error bars represent one standard deviation, and $k_{Nyquist}/2$ is indicated by the vertical dotted line. The predictions of the SP (dashed line) and PD (dotted-dashed line) fits are nearly identical, and are in excellent agreement with the simulations for both the weaker modifications with $\Sigma_0=\pm 0.008$ and the strong ones with $\Sigma_0=\pm 0.016$. The linear power spectra, showing the differences in linear growth factor arising from the modifications are shown by the dotted lines.}
\end{figure*}

\begin{figure*}[h!]
\centerline{\hbox{\hspace{0.0in}
\scalebox{0.90}{\includegraphics{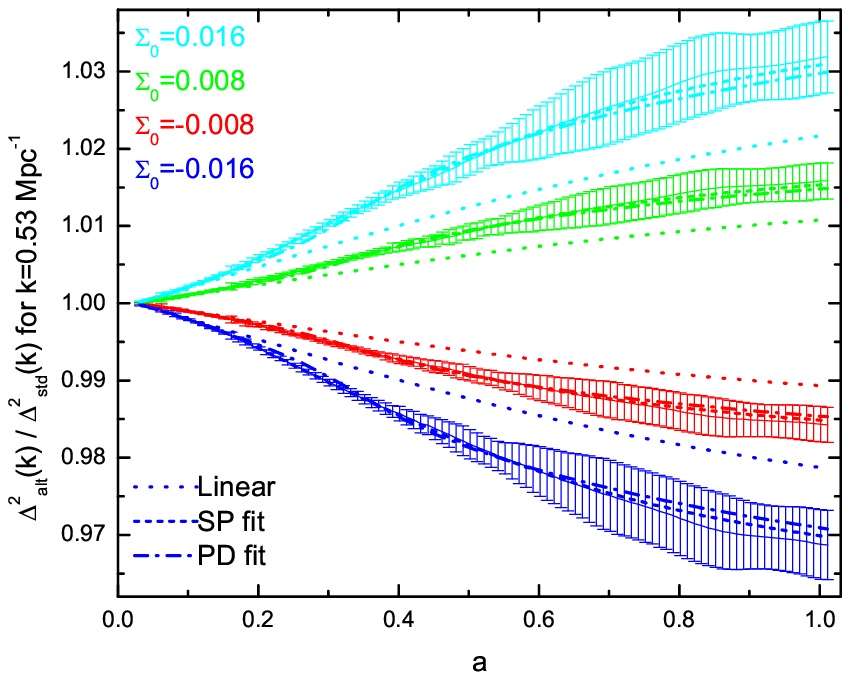}}
\scalebox{0.90}{\includegraphics{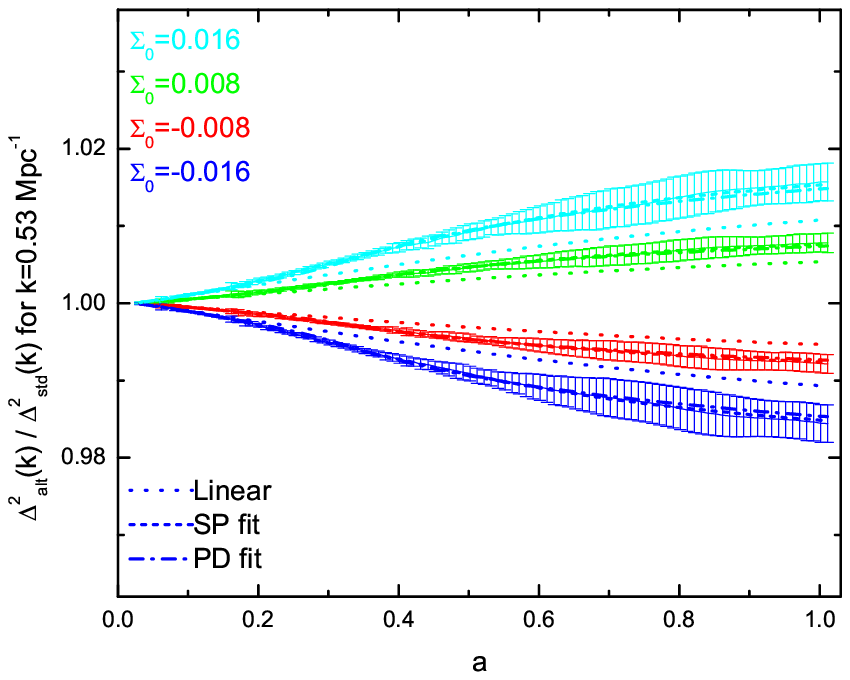}}}}
\caption{\label{fig:TMovera}
The evolution of the power spectrum over time for TM1 (left panel), and TM2 (right panel). Throughout the entire simulation the fits track the simulation results results extremely well. The color coding and line styles are the same as in Fig. \ref{fig:TMratios}.}
\end{figure*}

\begin{figure}[t!]
\scalebox{0.90}{\includegraphics{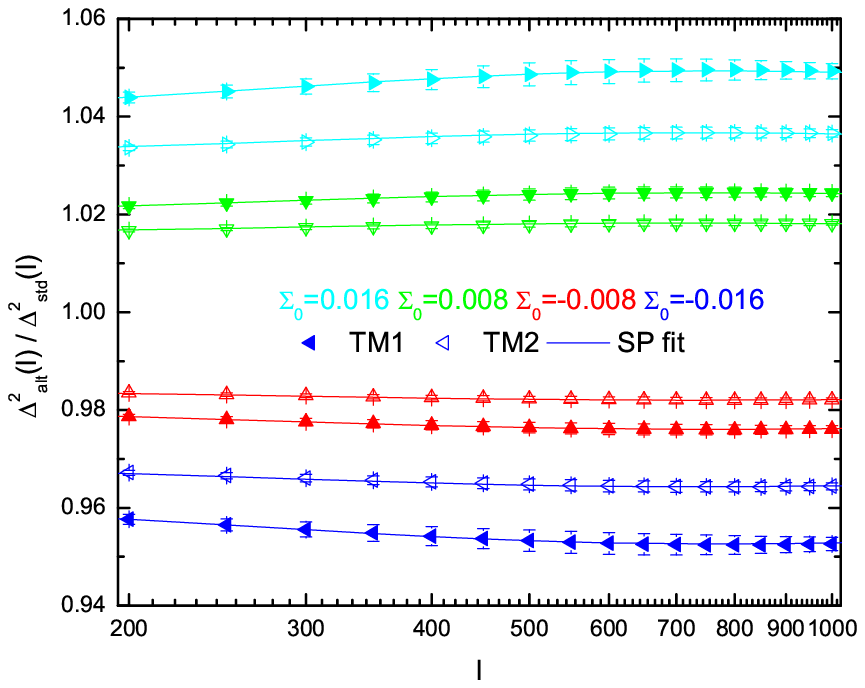}}
\caption{\label{fig:TM2WLPD24}
 The ratios of modified convergence power to standard convergence power in the twin models TM1 (full triangles) and TM2 (empty triangles) for $\Sigma_0 = 0.016$ ($\triangleleft$, blue), -0.008 ($\triangle$,red) ,0.008 ($\triangledown$, green), and 0.016 ($\triangleright$, light blue) shown against the predicted spectrum using the SP fit (full line), as the predictions of SP and PD are virtually identical.  As is to be expected, given the strong agreement between the fits and simulations of the matter power spectrum, the weak lensing spectra from the simulations are predicted well by the analytical fits.  }
\end{figure}

In order to investigate the abilities of the two analytical fits to predict nonlinear behavior in the two types of modifications, we consider a set of twin toy models, described in Sec. \ref{twin}.  TM1 has a modified Poisson equation $\{Q=1+\Sigma_0 a, \eta = 1\}$ while TM2 has anisotropic stress $\{Q=1,\eta=2\Sigma_0 a\}$. Both models have the same form of relationship of the weak lensing potential to the over-density, characterized by the function $\Sigma(k,a) = Q(1+\eta/2)$. As shown in Fig. \ref{fig:TM2WLPD24}, despite the degenerate background evolutions, the different modifications in each model lead to different linear scale-independent growth factors. For both models, the SP and PD analytical fits track both  the scale-dependent behavior and time evolution of nonlinearities in both types of scenario,  as shown in Figs. \ref{fig:TMratios} and \ref{fig:TMovera}. The weak lensing correlations for PD and SP fits are virtually identical for each model so we only show the results for SP fits  in Fig. \ref{fig:TM2WLPD24}; the difference between the simulations and analytical fits is negligible for both models.

\subsection{Discussion}

The nonlinear fits of Peacock and Dodds and Smith \textit{et al.} have been shown to work across broad cosmological models with standard gravity,  with different fractional mass densities, curvature, and initial power spectrum spectral indices. The utility of these fits derives from the wide applicability of the Zel'dovich approximation. In both fits there is the conjecture that the statistics of the gravitational clustering obey a similarity transform $P_NL(k/a) =\tilde{P}(k/k_{NL})$ for which no proof is given, but instead is experimentally shown to be robust for a variety of cosmological models by simulation. In this paper we assess whether such a similarity transform similarly exists in modified gravity theories, and moreover that the existing quantitative values for the fit coefficients can be used. This is not necessarily the case {\it a priori}.

 To test the fits we have performed nonlinear simulations of models in which modifications to Poisson and the peculiar acceleration occur exactly in this mildly nonlinear, transition regime. We have found that both the SP and PD analytical fits give reasonably good agreement with the simulations, in spite of the scale- and time-dependent modifications. This implies that applicability of the Zel'dovich approximation extends to scenarios in which anisotropic stress and even those with scale-dependent modifications to gravity are present in the mildly nonlinear regime. The modifications, therefore, are well described by the fits through their impact on the linear growth factor, $g$, and the spectral index dependency of the fitting functions. It appears that scale-dependent modifications in the mildly nonlinear regime do not require significant modification of the numerical coefficients in the fitting functions. Since our simulations focus on the ability of the fits to accurately match the transition from linear to nonlinear regimes, they do not investigate if modifications on small scales, in which the subhalo correlations are key, are well described by the fits, for example, if $r_s$ in (\ref{eq:UABeq}) were significantly smaller, e.g. less than $1$ Mpc. This is an area of interest for further analysis, especially in recently discussed theories in which galactic scale modifications could be present (e.g. \cite{Hu:2007nk}).

\section{Conclusions}\label{conc}
The use of complementary cosmological observations to probe the properties of dark energy  has proved extremely powerful. Observations sensitive to the background evolution, e.g. supernovae, or the wholly linear regime, e.g. the cosmic microwave background, have been the major observational tools to constrain dark energy to date. There is now significant interest, however, in applying a broader range of observations including those sensitive to large scale structure including large scale galaxy surveys, such as the Sloan Digital Sky Survey, and current and prospective weak lensing surveys. For each of these, in order to make precise inferences about dark energy, theoretical systematic errors about the modeling of nonlinear corrections must be addressed.

This work considers the ability  of  the commonly used nonlinear analytical fits of Peacock and Dodds \cite{Peacock:pd} and Smith \textit{et al.} \cite{Smith:sp} to predict nonlinear growth in a variety of theories beyond standard gravity. We consider models in which the Poisson equation is modified, based on 5D gravity,  \cite{Gregory:2000jc,Binetruy:2000xv,UAB:2001uab} and also those in which peculiar acceleration response to the gravitational potential is altered, including the DGP model  \cite{Dvali:2000hr}.  

We find that the two fitting functions provide robust predictions for theories with both types of modification, in terms of accurately predicting the matter power spectrum today, and also, vitally for calculating the weak lensing convergence spectrum, they predict the development of nonlinearities over time. Both consistently give predictions within 1$\sigma$ of 24 simulated N-body realizations of the theory. Our results imply that the similarity conjecture for mapping linear to nonlinear power empirically found to be satisfied in standard gravity simulations is also applicable to scenarios in which gravity has scale- and time-dependent modifications. This suggests that the spectral index dependence of the fitting function and the linear growth factor effectively describe alterations in the nonlinear collapse due to scale-dependent modifications to gravity and anisotropic stress at the scales studied in the models here.

We conclude that current analytic fits using the linear power spectrum in modified gravity theories can be used to accurately predict the nonlinear growth in theories with scale-independent or dependent modifications, and in those with or without anisotropic stress  in the mildly nonlinear regime. We find no statistical evidence for a preference, on the basis of overall performance, for one analytical fit over the other.  

Many modified gravity models, for example, DGP and $f(R)$ models, exhibit gravitational modifications on subhalo scales. Whether such modifications are well described by the halo term in the SP fit or the stable clustering approximation in the PD fit necessitates smaller scale simulations in the substantially nonlinear regime, which lies outside the scope of this paper.

We have limited our investigation of anisotropic stress to scale-independent modifications, and indeed further work is warranted in investigating whether the conclusions found for those are applicable to scale-dependent anisotropic stress, as found in $f(R)$ theories. It will also be interesting to investigate the agreement between  simulations and  the recently proposed nonlinear ansatz  \cite{Hu:2007pj} for modified gravity models. 


\section*{Acknowledgments}

We thank Anatoly Klypin and Jon Holtzman for kindly making their PM code publicly available and Olivier Dore, Hans Stabenau and Ira Wasserman for valuable discussions in the course of this work. The work of IL and RB is supported by the National Science Foundation under Grants No. AST-0607018 and No. PHY-0555216.
\clearpage
\bibliographystyle{apsrev}
\bibliography{paper}

\end{document}